\documentclass[aps,pra,superscriptaddress,10pt,notitlepage,longbibliography]{revtex4-1}
\usepackage{amsmath}
\usepackage{amsfonts}
\usepackage{amssymb}
\usepackage{graphicx}
\usepackage[utf8]{inputenc}
\usepackage[colorlinks=true,urlcolor=blue,linkcolor=red,citecolor=blue]{hyperref}

\usepackage{scalerel}
\usepackage{tikz}
\usetikzlibrary{svg.path}
\definecolor{orcidlogocol}{HTML}{A6CE39}
\tikzset{
  orcidlogo/.pic={
    \fill[orcidlogocol] svg{M256,128c0,70.7-57.3,128-128,128C57.3,256,0,198.7,0,128C0,57.3,57.3,0,128,0C198.7,0,256,57.3,256,128z};
    \fill[white] svg{M86.3,186.2H70.9V79.1h15.4v48.4V186.2z}
                 svg{M108.9,79.1h41.6c39.6,0,57,28.3,57,53.6c0,27.5-21.5,53.6-56.8,53.6h-41.8V79.1z M124.3,172.4h24.5c34.9,0,42.9-26.5,42.9-39.7c0-21.5-13.7-39.7-43.7-39.7h-23.7V172.4z}
                 svg{M88.7,56.8c0,5.5-4.5,10.1-10.1,10.1c-5.6,0-10.1-4.6-10.1-10.1c0-5.6,4.5-10.1,10.1-10.1C84.2,46.7,88.7,51.3,88.7,56.8z};
  }
}
\newcommand\orcid[1]{\href{https://orcid.org/#1}{\,\mbox{\scalerel*{
\begin{tikzpicture}[yscale=-1,transform shape]
\pic{orcidlogo};
\end{tikzpicture}
}{R}}}}

\newcommand{\ket}[1]{\left|#1\right\rangle}
\newcommand{\bra}[1]{\left\langle#1\right|}

\newcommand{\ketbra}[2]{\left|#1\right\rangle\left\langle#2\right|}

\newcommand{\w}{\omega}

\newcommand{\cL}{{\mathcal L}}
\newcommand{\cD}{{\mathcal D}}
\newcommand{\cC}{{\mathcal C}}
\renewcommand{\Re}{\text{Re}}
\renewcommand{\Im}{\text{Im}}
\newcommand{\mbar}{\bar{m}}
\newcommand{\Tr}[1]{\text{Tr}\left\{#1\right\}}

\newcommand{\ti}{\text{i}}
\newcommand{\td}{\text{d}}
\newcommand{\te}{\text{e}}

\makeindex

\begin{document}

\title{Breakdown signatures of the phenomenological Lindblad master equation in the strong optomechanical coupling regime}

\author{Ralf Betzholz\orcid{0000-0003-2570-7267}}
\affiliation{School of Physics, International Joint Laboratory on Quantum Sensing and Quantum Metrology,
Huazhong University of Science and Technology, Wuhan 430074, People's Republic of China.}
\author{Bruno G. Taketani\orcid{0000-0003-2741-8135}}
\affiliation{Departamento de Física, Universidade Federal de Santa Catarina, Florianópolis, 88040-900, Brazil.}
\author{Juan Mauricio Torres\orcid{0000-0002-7947-6962}}
\email{jmtorres@ifuap.buap.mx}
\affiliation{Instituto de F\'isica, Benem\'erita Universidad Aut\'onoma de Puebla, Apdo. Postal J-48,
Puebla 72570, Mexico.}

\begin{abstract}
The Lindblad form of the master equation has proven to be one of the most convenient ways to describe the impact of an environment interacting with a quantum system of interest. For single systems the jump operators characterizing these interactions usually take simple forms with a clear interpretation. However, for coupled systems these operators take significantly different forms and the full dynamics cannot be described by jump operators acting on the individual subsystems only. In this work, we investigate the differences between a common phenomenological model for the master equation and the more rigorous dressed-state master equation for optomechanical systems. We provide an analytical method to obtain the absorption spectrum of the system for both models and show the breakdown of the phenomenological model in both the bad cavity and the ultra-strong coupling limit. We present a careful discussion of the indirect dephasing of the optical cavity in both models and its role in the differences of their predicted absorption spectra. Our work provides a simple experimental test to determine whether the simpler phenomenological model can be used to describe the system and is a step forward toward a better understanding of the role of the coupling between subsystems for open-quantum-system dynamics.
\end{abstract}

\maketitle

\section{Introduction}
The theory of open quantum systems has been very  successful in describing the dynamics of many relevant physical settings taking into account the effects of the unavoidable coupling to their surroundings~\cite{Breuer,Carmichael,Carmichael1973}. Unlike closed systems, a proper description of open quantum systems can usually be attained in terms of a master equation (ME)~\cite{Carmichael} for the density operator of the system, where the environmental interaction leads to corrections to the unitary dynamics as well as non-unitary contributions to it. The nature of the relevant interactions and the regime of interest are determinant to the characterization and possible mathematical simplification of these contributions~\cite{Taketani2018}.

As the control of quantum systems gains importance in the development of new technologies, an accurate description of these dynamics becomes paramount. In general, there are two aspects to achieve such a correct description: a proper modeling of the physical system and its reliable mathematical characterization. For the former, a deep understanding of the underlying system is needed, which can be obtained theoretically, experimentally, or through simulations~\cite{Messinger2019,Puebla2019,Wang2019,Lambert2019}. The latter, on the other hand, is heavily dependent on the specific regime of parameters that characterize the system and the type of effects to be observed~\cite{Govia017,Schmit2020,Huybrechts2020}. 

Under the Born-Markov approximations, which assume a memory-less environment with no initial correlation with the system and a weak system-environment coupling, the ME can be cast in Lindblad form~\cite{Lindblad1976}. These assumptions often apply to quantum-optical laboratory settings as well as typical operating regimes for quantum computation and quantum simulation \cite{Nielsen2011}. Their precise mathematical treatment is thus of crucial and timely importance. In its diagonal form, the Lindblad ME is written in terms of Lindblad jump operators, each with an associated decay rate. For single quantum systems, these jump operators often have a very intuitive form. For instance, for interactions inducing energy decay, the jump operators annihilate energy quanta \cite{Stenholm1986,Briegel1993}. This ease to interpret the jump operators has lead to a common use of phenomenological models to describe more complex, multipartite quantum systems \cite{Briegel1993,Wilson-Rae2008b,Wang2015,Criger2016}. However, while these phenomenological master equations (PhMEs) have proven useful and give a reliable description of systems in several regimes of interest, they may lead to wrong or even unphysical predictions outside their scope of validity~\cite{Scala2007,Gonzalez2017}.  

As discussed in Sec.~\ref{sec:me}, jump operators of multipartite quantum systems usually act non-trivially on more than one subsystem and this effect becomes more relevant with increasing coupling~\cite{Scala2007,Hu2015}. This explains why for most cases the PhME is appropriate, and points to the necessity of more rigorous approaches as one reaches higher couplings in near-term experiments. Such an approach has already been developed for the paradigmatic Jaynes-Cummings model with cavity losses~\cite{Scala2007} and more recently for the optmechanical system with the introduction of the optmoechanical dressed-state master equation (DSME)~\cite{Hu2015}. A relevant point in both cases is the emergence of Lindblad operators describing combined decay channels, which is a manifestation of the coupling between the subsystem. While these models show significant differences, for optomechanical systems a similar method has been demonstrated to solve their eigenvalue problem in an exact form~\cite{Torres2019}.At first glance, it could appear as a more complicated equation, however, it has been shown that for the optomechanical system, the eigenvalue problem for both MEs, PhME and DSME, can be solved in an exact form~\cite{Torres2019}. This opens up new possibilities to analyze such systems in an analytical way.

In this work, we analyze the optomechanical system and propose a simple experimental measurement to witness the breakdown of the PhME. This system allows us to leverage the recently introduced technique to fully diagonalize the generator of both MEs~\cite{Torres2019} and thereby analytically evaluate its absorption spectrum. Using the spectral decomposition~\cite{Torres2019} of both MEs~\cite{Hu2015}, we analytically evaluate the cavity absorption spectrum. While the discussion above is in general valid for all interacting multipartite systems, optomechanical devices present two interesting advantages. First, experimentally these systems are nearing the parameter regime where the breakdown is expected to happen. Second, on the theory front, recently proposed techniques~\cite{Torres2019} can be leveraged to fully diagonalize the generator in a more rigorous ME than the phenomenological one, and thereby obtain the analytical form of the absorption spectrum. We show that in both the low-quality mechanical-oscillator and the ultrastrong-coupling regimes, the absorption spectra predicted by these two ME approaches are markedly different and that this difference is sensitive to changes in these parameters.The method presented here is not limited to optomechanical systems and can be adapted to other systems. We remark that the limits to PhME models have already been discussed in the context of the Jaynes-Cummings model~\cite{Scala2007,Beaudoin2011}, optomechanical systems~\cite{Hu2015,Naseem2018}, and in a general framework (we refer to Ref.~\cite{Carmichael1973} and references therein). The present work contributes to this discussion by proposing a simple experiment using standard setups in optomechanical devices, a complete analytical characterization of the expected measurement result for different ME models, as well as a better understanding of the regimes of applicability of the PhME.

This work is organized as follows: In Sec.~\ref{sec:me} we review two different ME approaches to describe interacting multipartite systems. Section~\ref{sec:optomechanics} introduces these two types of MEs for an optomechanical setup and analyses important differences. In Sec.~\ref{sec:absorption} we employ the diagonalization of the Liouville superoperator to derive the analytical form of the cavity absorption spectrum and discuss the emergence of discrepancies between the two MEs. In Sec.~\ref{sec:conclusion} we draw the conclusions.

\section{The Lindblad master equation}
\label{sec:me}
Let us briefly review the general formalism to derive a Lindblad-form ME and the main assumptions to arrive at the PhME and the DSME (for a detailed derivation we refer the reader to Refs.~\cite{Scala2007,Hu2015}). Consider a multipartite, interacting quantum system $S$ of interest, weakly coupled to an external environment. Without loss of generality, the system-bath interaction can be written as
\begin{align}
H_I=\sum_\alpha A_\alpha B_\alpha,
\end{align}
where $A_\alpha$ ($B_\alpha$) are Hermitian operators acting on the system (environment) Hilbert space only. Using the projection operators $\Pi(\epsilon)$ onto the system's eigenspace with energy $\epsilon$ we can define the operators
\begin{align}
A_\alpha(\w)=\sum_{\epsilon'-\epsilon=\hbar\w}\Pi(\epsilon)A_\alpha\Pi(\epsilon').
\label{eq:aalpha}
\end{align}
These operators induce transitions between system eigenstates with a fixed energy difference $\epsilon'-\epsilon$. After the Born-Markov and rotating-wave approximations, the non-unitary dynamics of $S$ can be written in the interaction picture with respect to its free Hamiltonian as
\begin{equation}
\label{eq:lindblad_general}
\mathcal{L}_I\rho=\sum_\w\sum_{\alpha,\beta}\frac{\tilde{\gamma}_{\alpha\beta}(\w)}{2}\left[2A_\beta(\w)\rho A_\alpha^\dagger(\w)-A_\alpha^\dagger(\w) A_\beta(\w)\rho-\rho A_\alpha^\dagger(\w) A_\beta(\w)\right],
\end{equation}
where $\mathcal{L}_I$ is a dissipation superoperator acting on the system density operator $\rho$ and $\tilde{\gamma}(\omega)$ are positive matrices containing information on the environment correlation functions~\cite{Breuer,Carmichael}. Diagonalizing these matrices and transforming into the Schrödinger picture leads to the usual Lindblad-form ME with the Liouville superoperator $\mathcal{L}$ given by
\begin{equation}
\label{eq:lindblad_diagonal}
\mathcal{L}\rho=
\frac{1}{\ti\hbar}[H,\rho]+
\sum_i\frac{\gamma_i}{2}\left(2L_i\rho L_i^\dagger-L_i^\dagger L_i\rho-\rho L_i^\dagger L_i\right).
\end{equation}
Here, $H$ is the free Hamiltonian of $S$ and $L_i$ are the Lindblad jump operators with their associated decay rates $\gamma_i$. The procedure outlined above is a general formalism to obtain a ME in the Born-Markov approximation.

For a multipartite system, the PhME approach approximates the full dissipator $\mathcal{L}_I$ by a sum of dissipators acting on the individual subsystems. However, if the subsystems are coupled, $\Pi(\epsilon)$ will project onto dressed states of the full system and therefore the operators $A_\alpha(\omega)$ will, in general, act non-trivially on all coupled subsystems. Thus, this approach is applicable to parameter regimes where the eigenstates of the full system can be well approximated by product states between the subsystem, such as for weakly coupled or far off-resonant systems.

Circumventing these limitations, the DSME incorporates the coupling between the subsystems by describing the full system and its interaction with the environment in its dressed-state basis (the eigenbasis of the coupled system, which, in general, does not consist of product states). Underlying this formalism are two assumptions on the environment. The first is that the subsystems couple to distinct, non-interacting environments. For optomechanical systems this is expected given the significant frequency differences between the subsystems. The second assumption is that the spectral density of one of the subsystem-environment interaction is flat around the central frequency of the corresponding subsystem. For the optomechanical case this subsystem is the optical cavity and the assumption is justified by the large frequency detuning to the mechanical mode. The cavity-bath contribution is then identical to that in the PhME~\cite{Hu2015}. However, other then an Ohmic spectral density, no assumption is made on the bath of the second subsystem, the mechanical oscillator. As discussed in Sec.
~\ref{sec:optomechanics}, this has two important effects. First, the jump operators appearing in the PhME are shifted proportionally to the cavity excitation number operator, with a proportionality factor dependent on the coupling strength. Unlike in the PhME, these parts of the dissipator will thus act non-trivially on both systems. The second effect is the presence of an additional cavity-dephasing term in the total dissipator, solely induced by the indirect interaction of the cavity with the thermal bath of the mechanical oscillator.

Importantly, the regime for which the DSME gives a reliable mathematical characterization of the system is not limited by the interaction strength, contrary to the PhME, but rather on the validity of the environmental assumptions discussed above. For optomechanical systems these are usually well justifiable and the dressed-state approach is therefore encouraged.
We note that these are distinct models, rather than the PhME being a limit to the more general DSME, and they only coincide in the zero-coupling limit.
Here, we use the optomechanical damping basis to fully diagonalize the Liouville superoperator for the optomechanical system~\cite{Torres2019} in both ME approaches, and use it to determine its absorption spectrum~\cite{Betzholz2014}. We remark that even for a weak coupling strength both the phenomenological and dressed-state approaches may lead to unphysical results such as the violation of the second law of thermodynamics~\cite{Naseem2018}.

\section{Master equations for optomechanical systems}
\label{sec:optomechanics}
In this section we present a standard optomechanical system~\cite{Aspemayer2014,Kippenberg2008} composed of a single-mode optical
cavity with one fixed and one movable mirror described by a mechanical oscillator. Both quantum mechanical degrees of freedom are coupled via the radiation pressure that is exerted by the photons of the optical cavity on the movable mirror. The Hamiltonian generating the unitary part of the dynamics is given by
\begin{equation}
H=\hbar\omega a^\dagger a+\hbar\nu b^\dagger b-\hbar\chi a^\dagger a(b+b^\dagger),
\label{Hamiltonian}
\end{equation}
which is written in terms of two pairs of bosonic operators, namely $a$ and $a^\dagger$
for the optical mode, and $b$ and $b^\dagger$ for the mechanical oscillator.
The first term in this Hamiltonian describes the energy of the optical cavity with frequency $\omega$, while the second one accounts for the harmonic motion of the mirror with frequency $\nu$. The last term describes the optomechanical interaction~\cite{Law1994,Law1995}, 
whose strength depends on the
photon number operator $a^\dagger a $ and the single-photon coupling strength $\chi$. In the absence of environmental interactions this Hamiltonian can be diagonalized using a photon-number dependent mechanical displacement transform~\cite{Hu2015}.

However, in a realistic scenario, the complete dynamics of the system cannot be faithfully accounted for without including loss effects in the system. These losses 
are the result of the interaction with the surroundings, which lead to photon
leakage from the cavity into a zero temperature bath and phononic damping
in the mechanical oscillator surrounded by a non-zero temperature bath.
The assumption of a zero temperature photonic bath is reasonable, since for optical frequencies and usual laboratory conditions the thermal photon number is negligible. Taking the effects of losses into account, a more complete description than the strictly unitary dynamics governed by $H$ is given by a Lindblad ME introduced in the previous section that can be written as $\partial\rho/\partial t=\cL\rho$,
where $\cL$ is the generator of the dynamics. The unitary and non-unitary parts of such a time evolution are conveniently written with the aid of the two superoperators 
\begin{equation}
\label{eq:superoperators}
\cC[X]\rho=[X,\rho],\qquad
\cD[X]\rho=2X\rho X^\dagger-X^\dagger X\rho-\rho X^\dagger X,
\end{equation} 
acting on the density matrix $\rho$, namely a commutator and a dissipation superoperator which was introduced in Sec.~\ref{sec:me}, respectively. In the remainder of this work, we will analyze the two forms of the Lindblad ME that were reviewed in the previous section for such an optomechanical system. The PhME is obtained by following the procedure outlined above assuming that the system Hamiltonian is merely $\hbar\omega a^\dagger a+\hbar\nu b^\dagger b$. The optomechanical interaction is simply added to the system Hamiltonian after this derivation. This results in a simple addition of two independent damping terms to the unitary dynamics and is the ME most commonly found in the literature
~\cite{Briegel1993,Wilson-Rae2008b,Wang2015,Criger2016}. Explicitly, the generator of the PhME has the form
\begin{equation}
\cL_{\rm Ph}=\frac{1}{\ti\hbar}\cC[H]
+\frac{\kappa}{2}\cD[a]
+\frac{\gamma(\mbar +1)}{2}\cD[b]+\frac{\gamma \mbar}{2}\cD[b^\dagger].
\label{masterPh}
\end{equation}
The first term in Eq.~\eqref{masterPh} describes the unitary dynamics
governed by the Hamiltonian in Eq.~\eqref{Hamiltonian} \footnote{We ignore small contributions from the Lamb-shift Hamiltonian as it does not change the form of Eq.(\ref{Hamiltonian}).}. The
 second term describes the losses from the optical mode in contact with a zero-temperature bath with decay rate $\kappa$, 
while the third and forth terms account for the damping of the mechanical oscillator with rate $\gamma$ by a heat bath at temperature $T$, resulting in a mean phonon number $\bar m=[\exp(\hbar\nu/k_\text{B}T)-1]^{-1}$.

For strong coupling between the two modes, however, the PhME is no longer appropriate. Especially in the ultra-strong coupling regime, $\chi/\nu\gtrsim 1$, a more accurate description is necessary. The reason is that the coupling also affects the decay mechanisms, as outlined in Sec.~\ref{sec:me} and evinced in Ref.~\cite{Hu2015}. Therefore, contrary to the PhME, the DSME is derived considering the full Hamiltonian, i.e., the same procedure is performed in the basis in which $H$, from Eq.~\eqref{Hamiltonian}, is diagonal.
This results in the generator~\cite{Hu2015}
\begin{equation}
\cL_{\rm DS}=\frac{1}{\ti\hbar}\cC[H]
+\frac{\kappa}{2}\cD[a]
+\Gamma\,
\cD[a^\dagger a]
+\frac{\gamma(\mbar +1)}{2}
\cD\left[b-\frac{\chi}{\nu}a^\dagger a\right]
+\frac{\gamma \mbar}{2}\cD\left[b^\dagger-\frac{\chi}{\nu}a^\dagger a\right].
\label{masterDS}
\end{equation}
There are two differences with respect to the PhME~\eqref{masterPh}. First, the third term 
describes a dephasing of the optical mode, which depends on the temperature of the mechanical oscillator's thermal environment, and was not present in the phenomenological model.
The dephasing decay rate is given by $\Gamma=2\gamma\chi^2/[\nu^2\ln(1+1/\mbar)]$. Later on, we will show that an analog parameter exists for the PhME, even when this dephasing process is not evident from the master  equation~\eqref{masterPh} itself. 
Second, the Lindblad jump operators in the dissipator of the mechanical mode are replaced according to $b\to b-\chi a^\dagger a/\nu$, leading to mixed decay channels in the system. Of course, in the weak-coupling and low-temperature limit both descriptions show similar behavior $\cL_{\rm Ph}\sim \cL_{\rm DS}$ as will be evidenced in Sec.~\ref{sec:absorption}. 

A remarkable feature of these models is that both MEs can be exactly solved, as was shown in Ref.~\cite{Torres2019}, where the eigensystem of both generators, $\cL_{\rm Ph}$ and $\cL_{\rm DS}$, was derived. 
Indeed one can solve the eigenvalue problem $\cL\hat\rho_\lambda=\lambda\hat\rho_\lambda$ and its dual problem, $\mathcal{L}^\dagger\check{\rho}_\lambda=\lambda^\ast\check{\rho}_\lambda$ for the non-Hermitian generators $\cL_{\rm Ph}$ and $\cL_{\rm DS}$~\cite{Briegel1993,Barnett2000}. Here, the set of the so-called right ($\hat\rho_\lambda$) and left eigenvectors ($\check{\rho}_\lambda$) is constructed to fulfill a completeness relation as well as the orthonormality relation $\text{Tr}\{\check\rho_\lambda^\dagger\hat\rho_\lambda\}=\delta_{\lambda,\lambda'}$ with respect to the Hilbert-Schmidt inner product.
Each eigenvalue (and thereby also the eigenvectors) depends on four integer parameters $l,k\in\mathbb{Z}$ and $n,m\in\mathbb{N}$ in the following form
\begin{gather}
\label{eq:eigenvalues}
\lambda\equiv
\lambda_{k,m}^{(l,n)}=-\ti l\omega-\frac{1}{2}(2n+|l|)\kappa-\ti k\nu-\frac{1}{2}(2m+|k|)\gamma
+\ti l|\beta|^2(2n+|l|)\nu-
l^2\Gamma.
\end{gather}
The first four terms correspond to the sum of eigenvalues from separate Liouvillians of two non-interacting damped harmonic oscillators, one for the optical and another for the mechanical mode. The contributions including the factors of $\beta$ and $\Gamma$, i.e., the last two terms, account for the coupling between the modes. The eigenvalues in Eq.~\eqref{eq:eigenvalues} hold for both the PhME and the DSME, and the only difference lies in the definition of the constants
\begin{equation}
\label{eq:betaGamma}
\beta=
\begin{cases}
\displaystyle\frac{\chi}{\nu-\ti\gamma/2},&{\rm PhME}\\
\\
\displaystyle\frac{\chi}{\nu},&{\rm DSME}
\end{cases},\qquad
\Gamma=
\begin{cases}
\left(\mbar+\displaystyle\frac{1}{2}\right)|\beta|^2\gamma,&{\rm PhME}\\
\\
2\left[\ln\left(\displaystyle\frac{\mbar+1}{\mbar}\right)\right]^{-1}\beta^2\gamma ,&{\rm DSME.}
\end{cases}
\end{equation}

\begin{figure}[tb]
\includegraphics[width=0.63\textwidth]{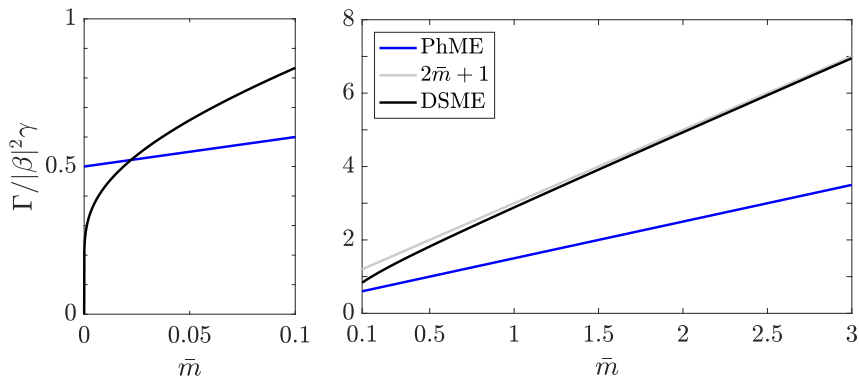}
\caption{\label{fig:Fig1} Dimensionless dephasing rate $\Gamma/|\beta|^2\gamma$ as a function of the mean thermal phonon number $\mbar$ for the two ME approaches, blue for the PhME and black for the DSME. The left panel shows the low temperature range and the right one higher temperatures. The gray line indicates the asymptotic value for the DSME, which is exactly twice as large as the value for the PhME.}
\end{figure}

The form of the eigenvalues sheds light on the characteristics  of both models. While it is evident from Eq.~\eqref{masterDS} that $\Gamma$ corresponds to the rate of a dephasing process in the DSME, there is no clear indication of pure dephasing in Eq.~\eqref{masterPh}. Although the presence of pure dephasing in the PhME can be seen in the interaction picture~\cite{Hu2015}, one can realize this by inspecting the analytical form of the eigenvalues in Eq.~\eqref{eq:eigenvalues}. In each eigenvalue $\lambda_{k,m}^{(l,n)}$, the factor $l^2$ accompanies $\Gamma$, where the integer $l$ labels the position of a diagonal in the optical-oscillator number space, while $n$ indicates the position on that diagonal (see Eqs.
~\eqref{eq:righteigenLEx} and \eqref{eq:lefteigenLEx} for the $n=0$ case). That is the reason for $n$ being positive, while $l$ can be any integer, positive or negative. Therefore $l\neq0$ indicates off-diagonal elements and $-l^2\Gamma$ expresses their corresponding irreversible decrease. Naturally, the damping mechanism with rate $\kappa$ also leads to a dephasing with rate $\kappa/2$. However, the form of $\Gamma$ in Eq.~\eqref{eq:betaGamma} is independent of $\kappa$ and consequently this is a distinct, pure dephasing mechanism not clearly seen in Eq.~\eqref{masterPh}. Another important feature is the dependence of $\Gamma$ on the mean phonon number $\mbar$. In order to analyze this dependence, in Fig.~\ref{fig:Fig1} we plot the dimensionless dephasing rate $\Gamma/|\beta|^2\gamma$ as a function of $\mbar$. 
For the PhME, the dependence on $\mbar$ is linear and does not vanish for $\mbar=0$, where it attains the value $1/2$. For the DSME, in contrast, the behaviour is linear only asymptotically for large values of $\mbar$, with the asymptotics $2\mbar+1$, i.e., it reaches values twice as large as in the PhME case. For low temperatures of the phononic bath, viz. for $\mbar \to 0$, one finds $\Gamma\to0$ for the DSME. This means that dephasing is twice as fast in the DSME for large values of $\mbar$, but ceases to exist in the limit of $\mbar\to 0$. This reinforces that these ME models are only equivalent for non-interacting systems as in both low and high-temperature limits the predicted pure dephasing rate differs. As will be seen in Sec.~\ref{sec:spectra}, these differences will play an important part in the differences of the absorption spectra predicted by both models.

In order to determine the absorption spectrum we will require the explicit form of the eigenvectors. The general exact analytical form of the full eigenbasis is shown in appendix~\ref{app:eig} or in Ref.~\cite{Torres2019}. Here we simply show the case $n=0$ and $l\ge 0$, which is the only subset of the eigenbasis required for the calculations in the next section. The right eigenvectors in this case read
\begin{equation}
\label{eq:righteigenLEx}
\hat{\rho}_{k,m}^{(l,0)}=|l\rangle\langle 0\vert\,
D^\dagger(\alpha_{l,0})\te^{-\eta_lb}\hat{\mu}_{k,m}\te^{\eta_lb}D(\beta_{l,0}),\quad
\end{equation}
while the left eigenvectors are given by
\begin{equation}
\label{eq:lefteigenLEx}
\check{\rho}_{k,m}^{(l,0)}=
|l\rangle\langle 0\vert\,
D^\dagger(\alpha_{l,0})\te^{-\eta_lb^\dagger}\check{\mu}_{k,m}\te^{\eta_lb^\dagger}D(\beta_{l,0})+\Upsilon,
\end{equation}
where $D(\alpha)=\exp(\alpha b^\dagger-\alpha^\ast b)$ is the displacement operator of the mechanical oscillator and we have used the three shorthand notations
$\alpha_{l,0}=-l\beta+l\mbar (\beta^\ast-\beta)$,
$\beta_{l,0}=l(\mbar+1) (\beta^\ast-\beta)$, and $\eta_l=l(2\mbar+1)(\beta-\beta^\ast)$. We note that for the DSME, where $\beta$ is real, these eigenvectors simplify remarkably, due to the fact that $\beta_{l,0}=\eta_{l,0}=0$ and $\alpha_{l,0}=-l\beta$. The last term in Eq.~\eqref{eq:lefteigenLEx}, $\Upsilon=\sum_{j=1}^{\infty}|j+l\rangle\langle j\vert\tilde\varrho_j$, does not contribute in the subsequent calculations of this work and the explicit form of the $\tilde\varrho_j$ can be found in appendix~\ref{app:eig}. Furthermore, $\hat\mu_{k,m}$ and $\check\mu_{k,m}$ are the right and left eigenvectors of a damped harmonic oscillator, whose forms are also given in appendix~\ref{app:eig}. 
From the form of the eigenvectors in Eqs.~\eqref{eq:righteigenLEx} and~\eqref{eq:lefteigenLEx} it is clear that,
as noted before, the integer index $l$ labels diagonals in the space of the optical mode, giving the dephasing interpretation to the $-l^2\Gamma$ term in Eq.~\eqref{eq:eigenvalues}.

In the following, we will show that the difference in the eigensystem of the two
MEs, solely characterized by $\beta$ and $\Gamma$, is sufficient in order to display different behaviors measureable in an absorption experiment.

\section{Signatures in the absorption spectrum}
\label{sec:absorption}
The absorption spectrum~\cite{Mollow1972,Cohen_Tannoudji1977,Betzholz2014} of the optical cavity is a convenient tool to query the system. For a weak probe laser of frequency $\omega_\text{p}$ driving the cavity, this spectrum is given by the Fourier transform of a two-time correlation function, evaluated in the steady state $\rho_\text{st}$, and has the form
\begin{equation}
\label{eq:absorption_1}
 {\mathcal A}(\omega_\text{p}) =  \frac{1}{2}\int\limits_{-\infty}^\infty \td t\, \langle [a(0),a^\dagger(t)]\rangle \te^{-\ti\omega_\text{p} t}.
\end{equation}
We will employ the spectral decomposition
of the Liouville superoperators in order to evaluate the lineshape of this spectrum.
The details of the analytical derivation are presented in Sec.~\ref{sec:spectra_evaluation}. 
However, the reader might skip it without loss of continuity and go to Sec.~\ref{sec:spectra}, where the final results are presented and discussed.

\subsection{Evaluation of the absorption spectrum}
\label{sec:spectra_evaluation} 
A great benefit of having the solution to the eigenvalue problem of a Liouville superoperator $\cL$ at hand is the possibility of finding the time evolution of any given initial condition $\rho_0$ as 
$\te^{\cL t}\rho_0=\sum_\lambda{\rm Tr}\{\check\rho_\lambda^\dagger\rho_0\}\te^{\lambda t}
\hat \rho_\lambda$~\cite{Briegel1993,Briegel1995}. This decomposition is also useful in the evaluation
of Eq.~\eqref{eq:absorption_1}  with the aid of the quantum regression theorem~\cite{Carmichael}. This is done by expressing the time dependence of the operator $a^\dagger(t)$ in terms of the  propagator $\exp(\cL t)$, which leads to
\begin{equation}
\label{eq:absorption_2}
 {\mathcal A}(\omega_\text{p})
 =\Re\int\limits_0^\infty {\rm d}t\,\Tr{a^\dagger \te^{\cL t}(\rho_\text{st} a)} \te^{-\ti\omega_\text{p} t}
 = 
  \Re\sum_{\lambda}\frac{\Tr{a^\dagger\hat\rho_\lambda}\Tr{\check\rho_{\lambda}^\dagger\ketbra{0}{1}\hat\mu_{0,0}}}{\ti\omega_\text{p}-\lambda}.
\end{equation}
Here, we have taken into account that  one of the two terms in the commutator in the correlation function vanishes, as the steady state of the system is given by
$\rho_\text{st}=\hat\rho^{(0,0)}_{0,0}=\ketbra{0}{0}\hat\mu_{0,0}$ 
(see Eqs. \eqref{righteigenL} and \eqref{eq:eigenvectorsDHO}, or Ref.~\cite{Bernad2015}).
The operator $\hat\mu_{0,0}$ is nothing but the thermal-equilibrium density operator of the mechanical oscillator, viz.  $[\mbar/(\mbar+1)]^{b^\dagger b}/(\mbar+1)$. 
In the last step above, the integral has been performed  using the fact that the real part of every non-zero eigenvalue is strictly negative.

Due to the simple form of the steady state it is possible to identify, by evaluating the second trace in the absorption spectrum of Eq.~\eqref{eq:absorption_2}, that the sum over eigenvalues $\lambda\equiv\lambda_{k,m}^{(l,n)}$ will be restricted to the indices $l=-1$ and $n=0$. Using the relations $\check\rho_{k,m}^{\dagger(-l,n)}=\check\rho_{-k,m}^{(l,n)}$ and $\check{\mu}_{k,m}^\dagger=\check{\mu}_{-k,m}$ together with Eq. \eqref{eq:lefteigenLEx} leads to the following expression
\begin{equation}
\label{eq:traceK}
    K_{k,m}=
\Tr{\check\rho_{k,m}^{\dagger(-1,0)}\ketbra{0}{1}\hat\mu_{0,0}}
=\Tr{\check{\mu}_{k,m}^\dagger \te^{\eta_1b^\dagger}D(\beta_{1,0})
	\hat\mu_{0,0} D^\dagger(\alpha_{1,0})\te^{-\eta_1b^\dagger}},
\end{equation}
where we have used the eigenvectors \eqref{eq:lefteigenLEx} and the cyclic property of the trace.  
The trace after the last equality sign in Eq. \eqref{eq:traceK}  can be carried out in terms of phase-space distributions according to the integral~\cite{Cahill1969a,Carmichael}
\begin{equation}
\label{eq:trace_PQ}
	K_{k,m}=\pi\int \td^2\alpha\, P_{k,m}(\alpha)Q_0(\alpha),
\end{equation}
where $P_{k,m}(\alpha)$ is the Glauber-Sudarshan $P$-distribution~\cite{Cahill1969b} of the left eigenvector
$\check\mu_{k,m}^\dagger=\check\mu_{-k,m}$ of the damped harmonic oscillator,
that is easily obtained from 
the antinormally-ordered eigenvectors $\check\mu_{k,m}$, 
 as written in Eq.~\eqref{eq:eigenvectorsDHO}. By
 simply replacing the annihilation and creation operators with the phase-space variables $\alpha$ and $\alpha^\ast$, respectively, one is able to find
\begin{equation}
P_{m,k}(\alpha)=\frac{(-1)^mm!}{\pi(m+|k|)!}
\alpha^{\frac{|k|+k}{2}}
\alpha^{\ast\frac{|k|-k}{2}}
L_m^{|k|}\left(\frac{|\alpha|^2}{\bar m +1}\right).
\end{equation}
On the other hand, the Husimi $Q$-function~\cite{Husimi1940} in Eq.~\eqref{eq:trace_PQ} is given by the expectation value of the remaining terms of $K_{k,m}$ in the coherent state $\ket\alpha$ and reads
\begin{equation}
Q_0(\alpha)=\frac{1}{\pi}\bra\alpha \te^{\eta_1b^\dagger}D(\beta_{1,0})
\hat\mu_{0,0}D^\dagger(\alpha_{1,0}) \te^{-\eta_1b^\dagger}\ket\alpha
=\frac{\te^{\ti(\bar m+1)\Im\{\beta^2\}-\frac{|\beta|^2}{2}}}{\pi(\bar m+1)}
    \te^{-\frac{|\alpha|^2}{\bar m+1}}\te^{
    \frac{\bar m}{\bar m+1}\alpha^\ast\beta
    -\alpha\beta^\ast}.
\end{equation}
We have arrived to this form  using the fact that $b\ket\alpha=\alpha\ket\alpha$ and standard properties of displacement operators, such as their quasi closure under multiplication and the Baker-Campbell-Hausdorff formula, in order to apply the displacement operators to the coherent states $\ket\alpha$. Then, we employed the normally-ordered form of the thermal state $\hat{\mu}_{0,0}$~\cite{Cahill1969a} given in Eq.~\eqref{eq:eigenvectorsDHO}, and used the notation in Eq. \eqref{alfabetas} or after Eq.~\eqref{eq:lefteigenLEx}. 
Substituting these two phase-space distributions in Eq.~\eqref{eq:trace_PQ} results in an integral of the form of Eq.~\eqref{eq:integral3}, whose derivation is given in appendix~\ref{app:integral}. By simplifying the resulting expression one can thereby find 
\begin{equation}
 K_{k,m}=\frac{(-1)^m
 	\te^{\ti(\bar m+1)\Im\{\beta^2\}-|\beta|^2(\bar m+\frac{1}{2})}
 }{(m+|k|)!}(\bar m\beta)^\frac{|k|+k}{2}[-(\bar m+1)\beta^\ast]^\frac{|k|-k}{2}(\bar m|\beta|^2)^m.
\end{equation}

 With the previously identified conditions, $l=-1$ and $n=0$, it is now simpler to evaluate
 the first trace in Eq.~\eqref{eq:absorption_2}
 using Eq.~\eqref{eq:righteigenLEx}, 
 together with the expressions $\hat\rho_{-k,m}^{(-l,n)}=\hat\rho_{k,m}^{\dagger(l,n)}$
 and $\hat{\mu}_{-k,m}^\dagger=\hat{\mu}_{k,m}$,
 namely
 \begin{equation}
 J_{k,m}= 
  \Tr{a^\dagger\hat\rho_{k,m}^{(-1,0)}}=
\Tr{D^\dagger(\beta_{1,0})
 	\te^{-\eta_1b^\dagger}\hat{\mu}_{k,m}
 	\te^{\eta_1b^\dagger}D(\alpha_{1,0})}.
 \end{equation}
 In this case, one can use the cyclic property of the trace and again an integral over phase-space distributions, similar to Eq.~\eqref{eq:trace_PQ}, in order to arrive at
 \begin{equation}
 J_{k,m}
 =\Tr{D(-\beta^\ast)\hat{\mu}_{k,m}}
 =
 \pi\int \td^2\alpha\, P_{-\beta^\ast}(\alpha)Q_{k,m}(\alpha),
 \end{equation}
  where $P_{\xi}(\alpha)=\exp(|\xi|^2/2+\xi\alpha^\ast-\xi^\ast\alpha)$ denotes the $P$-distribution of a displacement operator $D(\xi)$, whereas $Q_{k,m}(\alpha)=\bra\alpha\hat\mu_{k,m}\ket\alpha/\pi$ is the Husimi $Q$-function
 of the right eigenvectors of a damped harmonic oscillator, which is easily derived from their normally-ordered form~\eqref{eq:eigenvectorsDHO}. Evaluating the 
integral over these two phase-space distributions leads to 
 \begin{equation}
 	J_{k,m}=\frac{(-1)^m \te^{(\bar m+\frac{1}{2})|\beta|^2}
 	 	\te^{-(2\bar m+1)\beta^2+\ti(\bar m+1)\Im\{\beta^{2}\}}}{m!}
 	\beta^{\frac{|k|+k}{2}}(-\beta^\ast)^{\frac{|k|-k}{2}}[(\bar m+1)|\beta|^2]^m, 
\end{equation}
where we have employed again the integral~\eqref{eq:integral3}.

We remark that this calculation holds for both models, PhME and DSME. This is due to the fact that both eigensystems have the exact same form, with the difference appearing only in the definition of the parameters $\beta$ and $\Gamma$ given in Eq.~\eqref{eq:betaGamma}.

\subsection{Analysis of the absorption spectra}
\label{sec:spectra} 
By substituting the weight factors $J_{k,m}$ and $K_{k,m}$ derived above into Eq.~\eqref{eq:absorption_2}
and using the expressions of the eigenvalues presented in Eq.~\eqref{eq:eigenvalues},
one finds that the cavity absorption spectrum takes the form
 \begin{equation}
 \label{eq:absorption_3}
 {\mathcal A}(\omega_\text{p}) = \Re\sum_{k,m}
\frac{
	[\bar m(\bar m+1)|\beta|^{4}]^m
}{m!(m+|k|)!}
\left(\frac{\beta}{\beta^\ast}\right)^k
 \frac{
\left[\left(\bar m+\delta_{|k|,-k}\right)|\beta|^2\right]^{|k|}
\te^{-\bar m(\beta^2+\beta^{\ast 2})-\beta^{\ast 2}} 
 }{\ti(\omega_\text{p}-\omega+|\beta|^2\nu+k\nu)+\frac{\kappa}{2}+\left(m+\frac{|k|}{2}\right)\gamma+\Gamma}.
 \end{equation}
We stress again the fact that this expression holds for both the PhME and the DSME and the main difference between the two approaches lies in the parameters $\beta$ and $\Gamma$ in Eq.~\eqref{eq:betaGamma}. Considering that in most scenarios the condition $\gamma\ll\nu$ is likely to be fulfilled, one can presume that only the slightly different definitions of $\beta$ will not amount to a significant discrepancy in the absorption lineshape between the two models. The opposite case, i.e., if $\gamma$ is at least of the same order as $\nu$, on the other hand, might not be consistent with the Born approximation, on which the derivation of both MEs crucially depends. The significant difference between the dephasing rate $\Gamma$ in both approaches becomes most apparent by remembering two temperature limits. First, at zero temperature, $\Gamma$ vanishes for the DSME, while it takes a finite value for the PhME. Second, even for moderate mean thermal occupations, e.g., $\mbar\geq 2$, $\Gamma$ is already very close to its asymptotic value for the DSME (see Fig.~\ref{fig:Fig1}), which is bigger by a factor of two than the one in the PhME. Therefore, it is reasonable to assume that $\Gamma$ will be the main source for the discrepancies we will observe in the absorption lineshapes.

The form of the spectrum above shows that it is composed of a series of Lorentzians and Fano profiles. However, in the DSME description we have a real parameter $\beta$ and we see that the Fano profiles thereby do not contribute, yielding  a spectrum strictly built up by a sum of Lorentzian lineshapes. Furthermore, we see that the spectral position of every Lorentzian is determined by the integer $k$, meaning that the different Lorentzian components are separated by multiples of the mechanical frequency $\nu$. This also implies that each value of the integer $m$ adds a Lorentzian with the same central frequency, thus having an influence only on the height and width of the resulting phononic sideband, but not on its spectral position.  
In the zero-temperature limit, $\mbar\to0$, merely the case $m=0$ does not vanish, while only values $k\leq 0$ contribute to the spectrum.

It is worth mentioning that without the analytical result a numerical evaluation of the absorption spectrum can already become demanding even for moderate temperatures, with mean phonon occupation close to $\mbar=30$. This makes a faithful full numerical treatment challenging for high temperatures, while our expression holds for arbitrary values of $\mbar$ in both master-equation approaches. However, even with our analytical result, the summation runs over an infinite number of values of $k$ and $m$. To overcome this problem, one can estimate cutoff values for $k$ and $m$ in the sum of Eq.~\eqref{eq:absorption_3} by comparing  the weight factors for each Lorentzian with Poissonian distributions with mean value $\sqrt{\mbar(\mbar+1)}|\beta|^2$ for $m$ and $|\beta|^2(\mbar+\delta_{|k|,\pm k})$ for $\pm k$. In each of these two distributions, for $k$ and $m$, the mean value and variance are smaller than for a Poissonian distribution. Therefore, it is possible to estimate the cutoff values to accurately approximate the absorption spectrum, leading to the conditions $m\le 2\sqrt{\mbar(\mbar+1)}|\beta|^2$ and $-3(\mbar+1)|\beta|^2\le k\le 3\mbar|\beta|^2$.

\begin{figure}[tb]
\begin{center}
(a)\hspace{2mm}\vtop{\vskip-0ex\hbox{\includegraphics[width=0.42\textwidth]{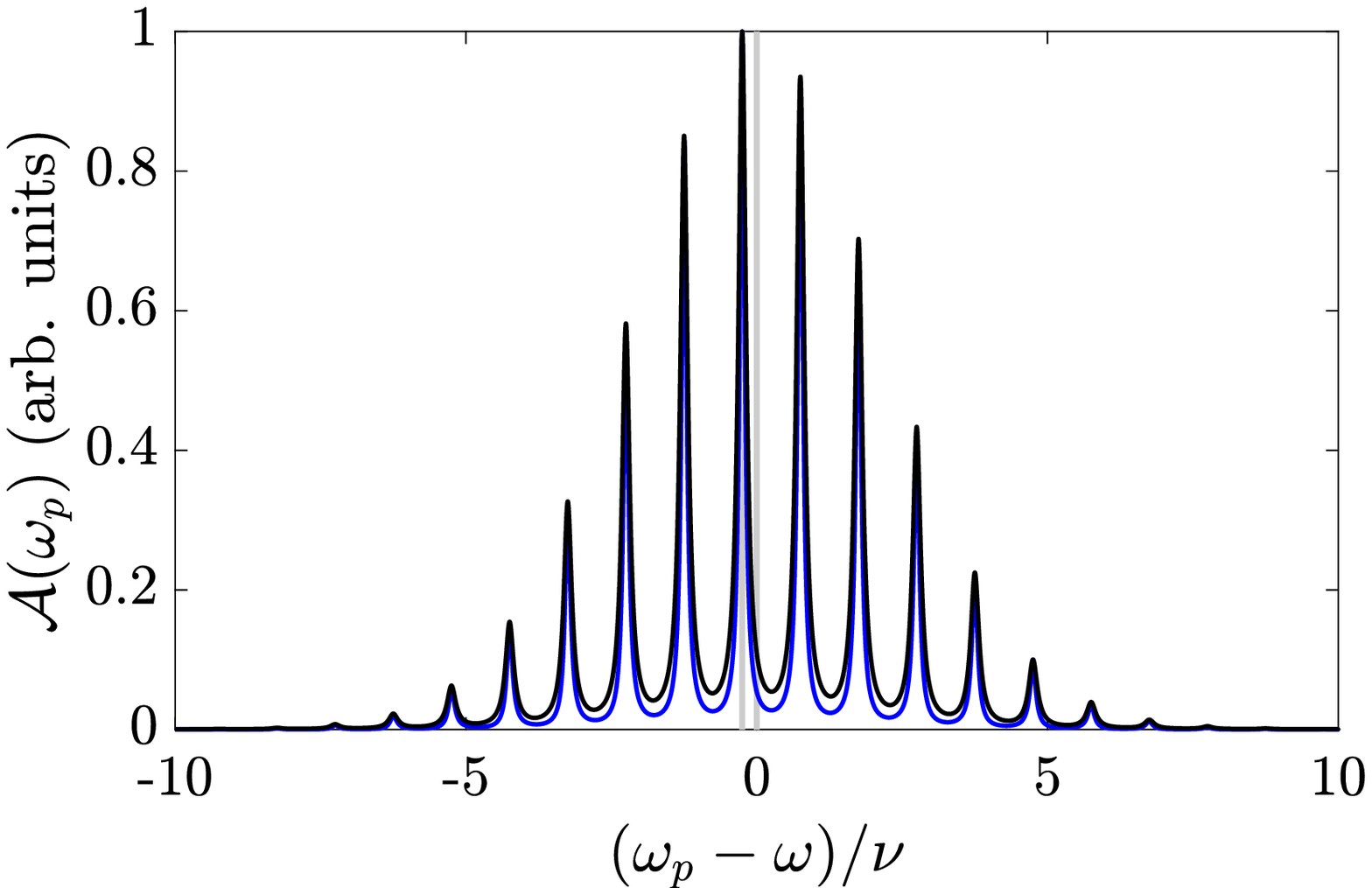}}}\hspace{8mm}
(b)\hspace{2mm}\vtop{\vskip-0ex\hbox{\includegraphics[width=0.42\textwidth]{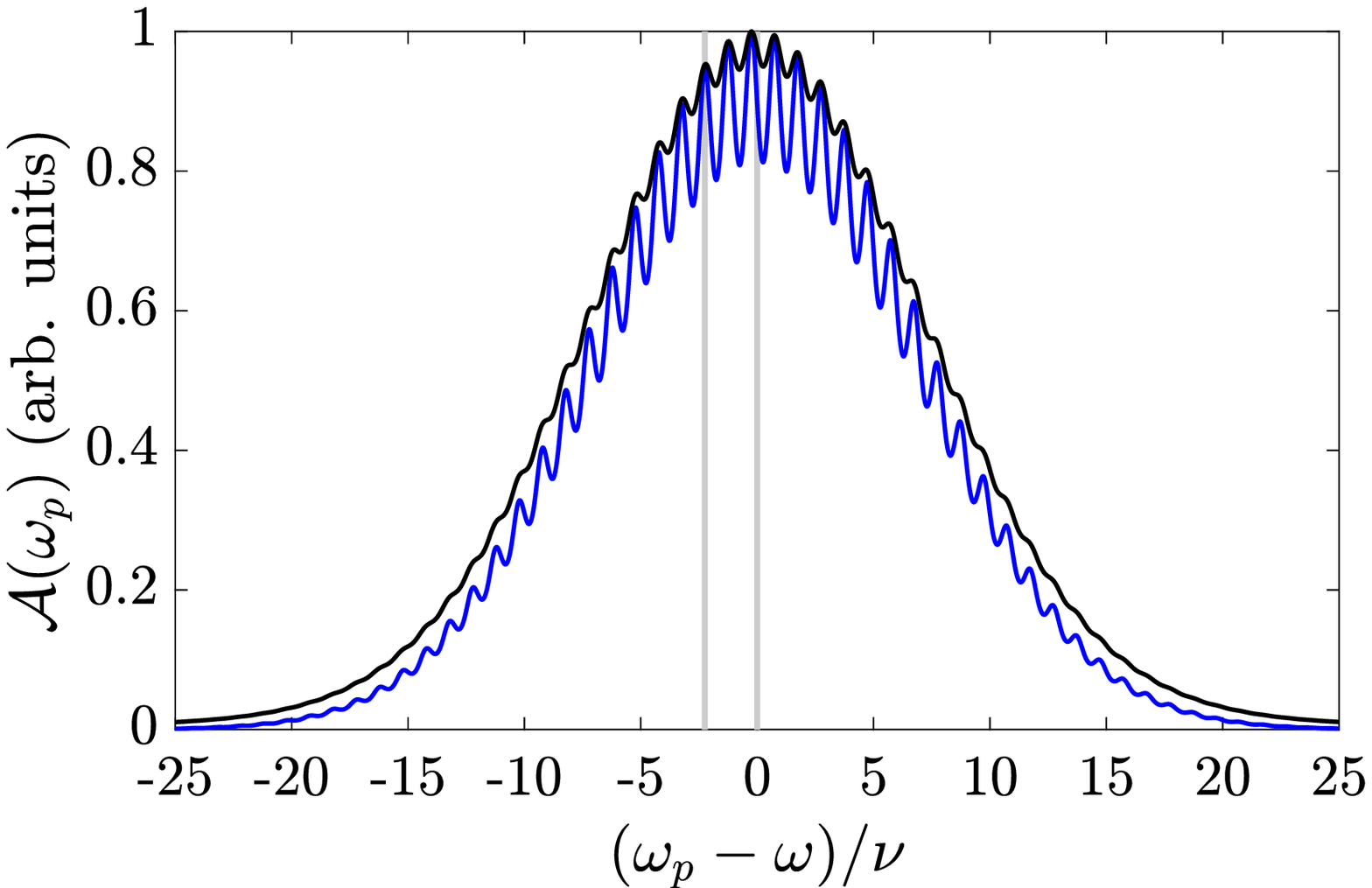}}}
\end{center}
\caption{\label{fig:Fig2} Cavity absorption spectra for different values of $\chi$: (a) $\chi=\nu/2$ and (b) $\chi=3\nu/2$. The remaining parameters are $\kappa=\gamma=\nu/100$ and $\mbar=10$. Blue lines show the PhME approach and black lines the DSME. Gray vertical lines show the position of the bare cavity resonance (right) and the zero-phonon line (left), i.e. the bare cavity resonance including the polaron shift $|\beta|^2\nu$.}
\end{figure}

In Fig.~\ref{fig:Fig2} we show the absorption spectra for two distinct values of the coupling strength $\chi$, namely for $\chi=\nu/2$ in Fig.~\ref{fig:Fig2}(a) and for $\chi=3\nu/2$ in Fig.~\ref{fig:Fig2}(b), both evaluated for the parameters $\kappa=\gamma=\nu/100$ and $\mbar=10$. 
Each spectrum itself is normalized such as to have a unit maximum. For $\chi=\nu/2$ we find that the two approaches have very good agreement, differing only by a small difference in contrast. Clear sidebands are observed as $\Gamma$ is significantly smaller than the distance of the peaks. For $\chi=3\nu/2$, on the other hand, we see a clear qualitative difference in the absorption lineshape. The larger coupling increases $\Gamma$ in both models broadening the Lorentzians and decreasing the contrast. However, as for these parameters $\Gamma$ is larger in the DSME, this case displays starker contrast loss.

\begin{figure}[tb]
\begin{center}
(a)\hspace{2mm}\vtop{\vskip-0ex\hbox{\includegraphics[width=0.42\textwidth]{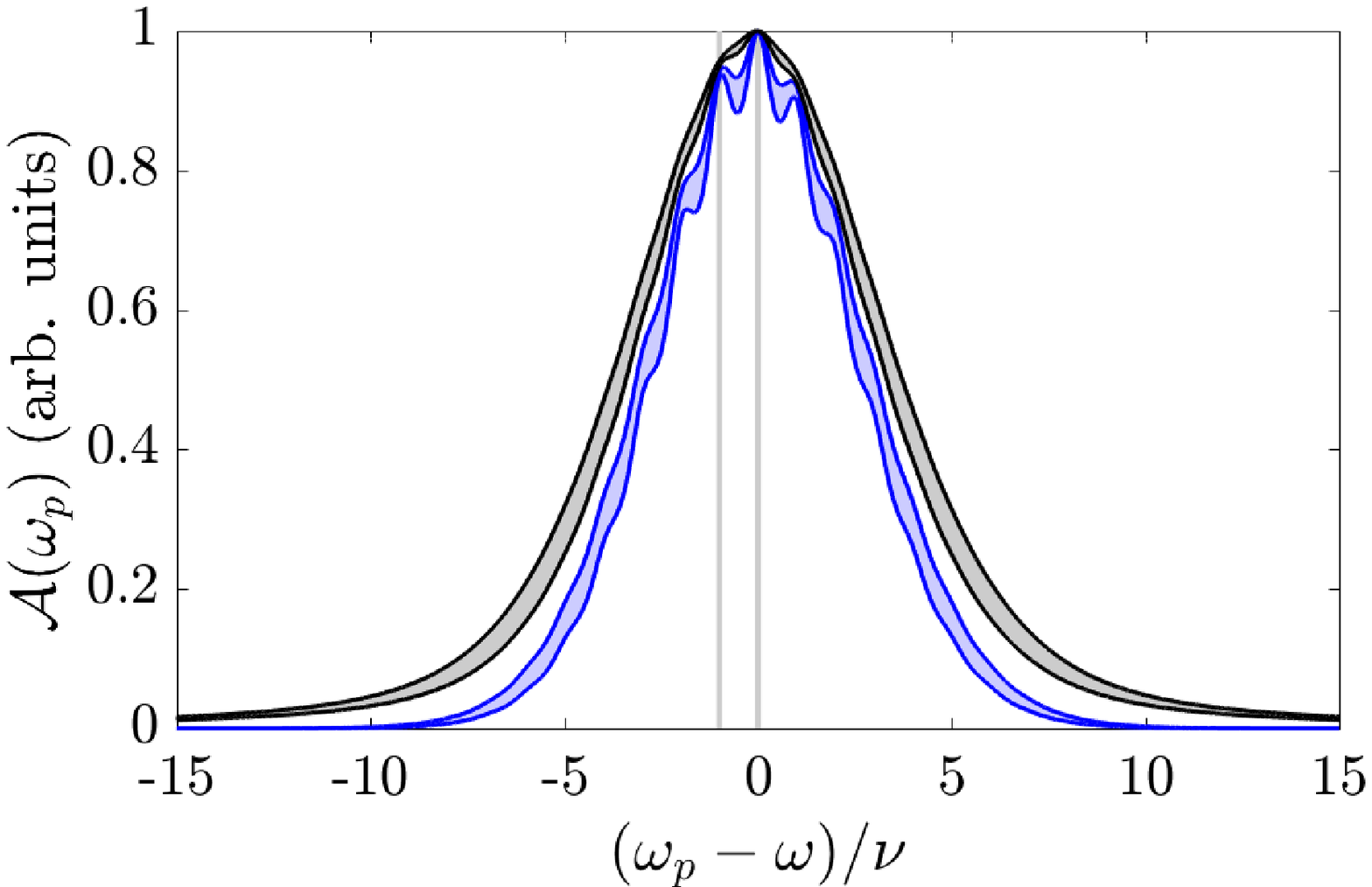}}}\hspace{8mm}
(b)\hspace{2mm}\vtop{\vskip-0ex\hbox{\includegraphics[width=0.42\textwidth]{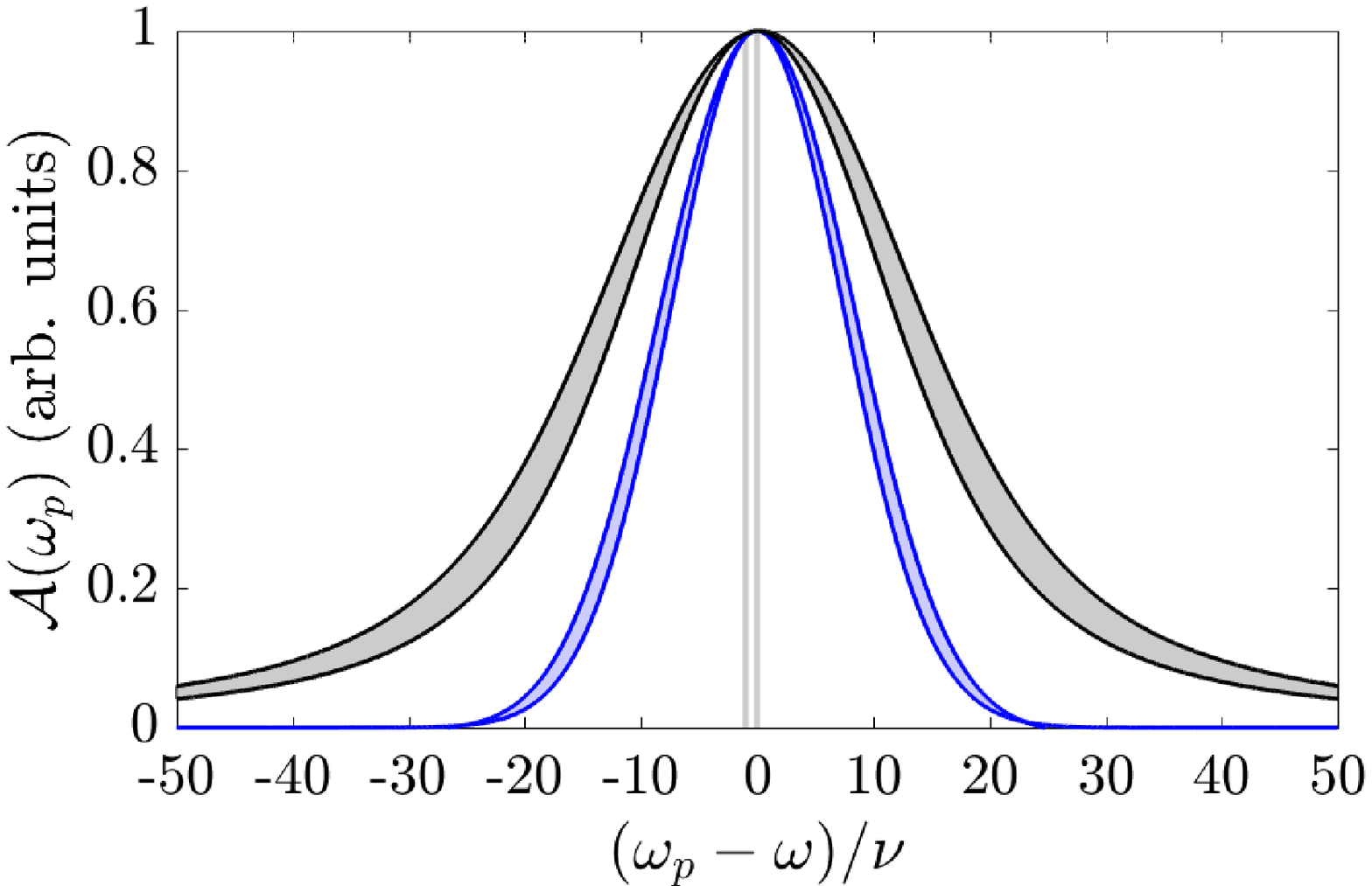}}}
\end{center}
\caption{\label{fig:Fig3} Cavity absorption spectra in the very low-$Q$ regime, with relatively pronounced differences in their shape: (a) $\mbar=3$ with already washed-out peaks in the DSME approach and (b) $\mbar=30$, where the full width at half maximum differs by roughly a factor of 1.7. The remaining parameters are $\chi=\nu$, $\kappa=\nu/100$, and $\gamma=\nu/10$. The shaded areas are outlined by $\mbar\pm\Delta\mbar$ and for both cases the uncertainty is $\Delta\mbar/\mbar=0.1$. Blue lines show the PhME approach and black lines the DSME. Gray vertical lines show the position of the bare cavity resonance (right) and the zero-phonon line (left).}
\end{figure}

In Fig.~\ref{fig:Fig3} we analyze the influence of the mechanical bath temperature by showing the absorption lineshape for two values of the thermal occupation, namely $\mbar=3$ in Fig.~\ref{fig:Fig3}(a) and $\mbar=30$ in Fig.~\ref{fig:Fig3}(b). 
Both cases are presented in the very low quality factor (low-$Q$) regime of the mechanical oscillator, viz. $\gamma=\nu/10$, and the other parameters are $\chi=\nu$ and $\kappa=\nu/100$. 
For both temperatures the two ME approaches are clearly distinguishable. For $\mbar=3$ the PhME spectrum still predicts peaks and dips, while in the DSME spectrum the separate spikes have already blended into a single broad peak. For $\mbar=30$ both approaches yield a spectrum consisting of an inhomogeneously-broadened single peak of Gaussian shape, whose full width at half maximum, however, differs by roughly a factor of 1.7 between the broad DSME spectrum and the narrower PhME spectrum.

It is interesting to notice that in both in the ultra-strong coupling and the low-$Q$ limits the predicted spectra is quite sensitive to small changes in the parameters, particularly for the DSME. This is advantageous as it allows one to use changes in the spectrum to determine the validity of the PhME for a given experimental situation.
In order to investigate how susceptible the lineshapes are for fluctuations in the mean thermal occupation, in every spectrum in Fig.~\ref{fig:Fig3}, we show shaded areas that are outlined by $\mbar\pm\Delta\mbar$, with a relative uncertainty $\Delta\mbar/\mbar=0.1$ in both cases. Even in the presence of such errors the predicted spectra are markedly distinct. Similar results were obtained for errors in the determination of $\chi$. These results indicate that the qualitative behaviour of the absorption spectrum of optomechanical systems can be used to probe whether a PhME as even in the presence of experimental uncertainties the predictions are significantly different.

\section{Conclusions}
\label{sec:conclusion}
With the experimental progress towards strongly coupled systems, a deeper understanding of their open dynamics is paramount. To this end, microscopic derivations~\cite{Scala2007,Hu2015,Naseem2018} of Lindblad MEs have been developed to improve the description over the simpler phenomenological approach. 
Focusing on the optomechanical system, this work builds on the recently introduced DSME~\cite{Hu2015} by describing a method that allows for the exact analytical determination of the absorption spectrum of the system. This, in turn, permits us to describe its behaviour outside the low temperature limit, where numerical methods are applicable. One important aspect to notice in the absorption spectrum is that important differences between the models only become significant in a few regimes, namely in the ultra-strong optomechanical coupling and the low-Q mechanical resonator regimes. Interestingly, in all cases, once the predictions start to differ, they actually become quite sensitive to changes in that parameter, be it the coupling or the quality factor. It is therefore very important to any model working close to these limits, to use the DSME as it gives a more complete description. 

Our work points to the absorption spectrum, a simple experiment, as a tool to determine whether a given experimental setup can be correctly described by the PhME. Figure~\eqref{fig:Fig3}, in particular, shows that the expected behaviour of the two approaches is significantly different, even within experimental uncertainties. From a theoretical point of view, it also aids in determining whether the DSME approach is needed, based on the parameter regimes one wants to explore. 

It is crucial to understand that the PhME and DSME approaches are not limits of some expansion of the full ME, but are rather distinct models to describe the system. They only match as the subsystems decouple and the dynamics can be effectively treated as that of separable systems. As we point out, these two models differ not only by the changes in decay mechanisms, but also in their effective dephasing mechanism (with rate $\Gamma$). This means that, as long as the subsystems are coupled, the two models predict different behaviour, regardless of the temperature regime. For moderate mean phonon occupation the DSME predicts a decay twice as large as that of the PhME. This rather interesting result is also seen in the plots of the absorption spectrum, where the visibility of the DSME reduces faster than in the PhME, since a higher dephasing rate translates to broader sideband linewidths in the spectrum. It is also to be noticed that while important, this does not account for all the changes in the predicted absorption spectra. 

Another important feature is that, in contrast to the case of the PhME, the absorption spectrum for the DSME is built up only in terms of Lorentzian profiles. This ensures that the spectrum will always be strictly positive, a condition that might be broken for certain parameters in the PhME. This important feature has its roots in the simpler form of the eigensystem as we have verified in appendix~\ref{app:eig} (and noted in Ref.~\cite{Torres2019}). 
This encourages the use of a DSME rather than a phenomenological description, as the work of a microscopic derivation is rewarded when solving and applying the eigenvalue problem. 

This result rests on the property that the optomechanical model is exactly solvable and on how the jump operators act on the eigenstates of the system. Analog behavior can be found in other quantum-optical systems~\cite{Torres2014}, so one can conjecture the same feature to hold for them, i.e., a simpler diagonalization of the Liouville superoperator for the corresponding DSME. It would be interesting to analyze this in the microscopic derivations of different quantum-optical models and to apply the  presented analytical machinery to them.

Our results clarify the scenarios where different descriptions can be applied. In the low temperature regime and low total excitation, numerical methods can be successfully used. Below the strong coupling regime and for mechanical quality factors above $100$, the PhME gives a reliable description. This also explains why this approach has found success and is widely adopted. Outside this regime the DSME is needed for a faithful description. It is important to notice that with increasing coupling and decay rates the assumptions on the DSME may also breakdown, including the Born-Markov approximation, which would then require non-Markovian ME models.

\section*{Acknowledgements}
R.B. would like to thank Meichen Yu for stimulating discussions during the early stages of this work and the National Natural Science Foundation of China for financial support under Grant No.~11950410494. B.G.T. acknowledges support from FAPESC, CNPq INCT-IQ (465469/2014-0) and Coordenação de Aperfeiçoamento de Pessoal de Nível Superior - Brasil (CAPES) - Finance Code~001. B.G.T. is grateful to Simon J\"ager for discussions and also to Murray Holland for their hospitality at JILA.
J.M.T. is grateful to Departamento de F\'isica, Federal University of Santa Catarina for the hospitality during his research stay and acknowledges support by SEP Project No.~PRODEP-511-6/18-9344.

\appendix
\section{Eigenvectors Liouville superoperator}
\label{app:eig}
Here, we present the full expressions of the  eigenvectors solving the DSME and the PhME. The form of the solutions is the
same for both systems, but depends on the parameter $\beta$ that has 
a particular form for each case, as given in Eq.~\eqref{eq:betaGamma}.
The right eigenvectors 
 \begin{equation}
  \label{righteigenL}
 \hat{\rho}_{k,m}^{(l,n)}=\sum_{j=0}^{n}|j+l\rangle\langle j\vert\hat{\varrho}_{k,m}^{(l,n;j)},
 \qquad
 \hat\varrho_{k,m}^{(l,n;j)}=
 \sqrt{\frac{n!(n+l)!}{j!(j+l)!}}
 {\sum^{n-1}_{\substack{\{k_r,m_r\}\\r=j}}}
 \Bigg[
 \prod_{s=j}^{n-1}
 \frac{\kappa c_{k_s,m_s}^{l,k_{s+1},m_{s+1}}}{\lambda_{k,m}^{(l,n)}-\lambda_{k_s,m_s}^{(l,s)}}\Bigg]\hat{\mu}_{k_{j},m_{j}}^{(l,j)},
 \end{equation}
  solve the eigenvalue equation 
  $\cL_{\rm Ph/DS}\hat\rho_{k,m}^{(l,n)}=\lambda_{k,m}^{(l,n)}\hat\rho_{k,m}^{(l,n)}$, with the eigenvalues $\lambda_{k,m}^{(l,n)}$
  given in Eq.~\eqref{eq:eigenvalues}. In this notation, the sum runs over the set of indices $\{k_r,m_r\}^{n-1}_{r=j}$, indicated in the lower and upper bound of the sum symbol. The left eigenvectors  
 \begin{equation}
 \label{lefteigenL}
 \check{\rho}_{k,m}^{(l,n)}=\sum_{j=n}^{\infty}|j+l\rangle\langle j\vert\check{\varrho}_{k,m}^{(l,n;j)},\qquad
 \check{\varrho}_{k,m}^{(l,n;j)}=
  \sqrt{\frac{n!(n+l)!}{j!(j+l)!}}
 {\sum^{j}_{\substack{\{k_r,m_r\}\\r=n+1}}}
 \left[\prod_{s=n+1}^{j}\frac{\kappa c_{k_{s-1},m_{s-1}}^{\ast l,k_s,m_s}}{\lambda_{k,m}^{\ast(l,n)}-\lambda_{k_s,m_s}^{\ast(l,s)}}\right]\check{\mu}_{k_j,m_j}^{(l,j)},
 \end{equation}
solve the dual eigenvalue equation 
 $\cL_{\rm Ph/DS}^\dagger\check\rho_{k,m}^{(l,n)}=\lambda_{k,m}^{\ast(l,n)}\check\rho_{k,m}^{(l,n)}$. The dual superoperator $\cL_{\rm Ph/DS}^\dagger$  has the same 
form as $\cL_{\rm Ph/DS}$ in Eqs.~\eqref{masterDS} and~\eqref{masterPh}, but  with 
$\cD^\dagger [X]\rho=2X^\dagger\rho X-X^\dagger X\rho-\rho X^\dagger X$ replacing
the dissipator $\cD [X]$. 

In the previous expressions for the eigenvectors, we have employed the displacement operators $D(\alpha)=\exp(\alpha b^\dagger-\alpha^\ast b)$ in the
following operators of the mechanical mode 
\begin{equation}
\label{eq:diplacedmu}
\hat{\mu}_{k,m}^{(l,n)}=D^\dagger(\alpha_{l,n})\te^{-\eta_lb}\hat{\mu}_{k,m}\te^{\eta_lb}D(\beta_{l,n}),
\qquad
\check{\mu}_{k,m}^{(l,n)}=D^\dagger(\alpha_{l,n})\te^{-\eta_lb^\dagger}\check{\mu}_{k,m}\te^{\eta_lb^\dagger}D(\beta_{l,n}),
\end{equation}
that can be thought of as asymmetrically displaced~\cite{Betzholz2014,Torres2019} right and left eigenvectors of a damped harmonic oscillator~\cite{Briegel1993,Englert2002}. We have chosen to represent these damped harmonic oscillator eigevectors in their
normally- and antinormally-ordered form as
\begin{equation}
\label{eq:eigenvectorsDHO}
\hat{\mu}_{k,m}=\frac{1}{(\mbar+1)^{k+1}}b^{\dagger k}\left\{L_m^{(k)}\left(\frac{b^\dagger b}{\mbar+1}\right)\te^{-\frac{b^\dagger b}{\mbar+1}}\right\}_{\rm n},
\qquad
\check{\mu}_{k,m}=\frac{m!}{(m+k)!}\left\{L_m^{(k)}\left(\frac{b^\dagger b}{\mbar+1}\right)\right\}_{\rm a}b^{\dagger k},
\end{equation}
respectively, with the generalized Laguerre polynomials $L_m^{(k)}(x)$. Here, $\{\,\cdot\,\}_\text{n}$ denotes normal and $\{\,\cdot\,\}_\text{a}$ antinormal ordering of the enclosed expression. The expressions in Eq.~\eqref{eq:eigenvectorsDHO} are only valid for positive $k$, while for negative values one has to take the Hermitian conjugate and replace $k$ by its absolute value. We have also introduced in Eqs.~\eqref{righteigenL} and~\eqref{lefteigenL} the following notation
\begin{equation}
c_{k',m'}^{l,k,m}=\Tr{\check{\mu}^{(l,n-1)\dagger}_{k',m'}\hat{\mu}^{(l,n)}_{k,m}}=
\te^{l(\beta^2-\beta^{\ast 2})/2}
\frac{
m'!|\beta|^{2(m'-m)+|k'|-|k|}
(\mbar+1)^{m-m'}e^{\ti\phi(k'-k)}
}{m!(m'-m-|k|\jmath_-)!(m'-m+|k'|-|k|\jmath_+)!},
\end{equation}
representing traces involving the operators shown in~\eqref{eq:diplacedmu}, where we defined  $\phi=\arg(-\beta)$ and the abbreviation $\jmath_\pm=|k/|k|\pm|k'|/k'|/2$. Furthermore, we have used in Eq.~\eqref{eq:diplacedmu} the displacement constants
\begin{equation}
\alpha_{l,n}=-(n+l)\beta+l\mbar (\beta^\ast-\beta),\qquad
\beta_{l,n}=-n\beta+l(\mbar+1) (\beta^\ast-\beta),\qquad \eta_l=l(2\mbar+1)(\beta-\beta^\ast),
\label{alfabetas}
\end{equation}
which simplify remarkably for the real $\beta$ found in the DSME case. Until now, all expressions in this appendix were only defined for positive values of the integer $l$. 
In order to complete the set of eigenvectors, 
the identities 
\begin{equation}
\hat\rho_{-k,m}^{(-l,n)}=\hat\rho_{k,m}^{\dagger(l,n)},\qquad
\check\rho_{-k,m}^{(-l,n)}=\check\rho_{k,m}^{\dagger(l,n)},\qquad
\lambda_{-k,m}^{(-l,n)}=\lambda_{k,m}^{\ast(l,n)}
\end{equation}
can be used to obtain the expressions for negative values of $l$.

\section{Useful integral}
\label{app:integral}
During the calculation of the absorption spectrum, in two separate instances we encountered complex integrals, that arose while carrying out traces over products of operators, and which we evaluated by integrating the product of their associated phase-space distributions. Both these integrals are special cases of the more general integral
\begin{equation}
\label{eq:integral1}
I_{\jmath,\ell}=\int\td^2\xi\,\xi^\ell  L_\jmath^{(\ell)}\big(q|\xi|^2\big)\te^{-c|\xi|^2}\te^{\varsigma\xi^\ast-\zeta^\ast\xi},
\end{equation}
where $\jmath$ and $\ell$ are positive integers  and $q$ and $c$ are constants with $\Re\,(c)>0$. For the sake of completeness and later reference, the derivation of this integral will be shown in this appendix.

A common way of calculating this kind of integrals is to employ the generating function of the Laguerre polynomials. However, here, we show a direct integration using polar coordinates $\xi=r\exp(\ti\varphi)$ for the complex plane, in which the integral takes the form
\begin{equation}
I_{\jmath,\ell}=\int_0^\infty\td r\,r^{\ell+1}L_\jmath^{(\ell)}\big(qr^2\big)\te^{-cr^2}\int_0^{2\pi}\td\varphi\,\te^{\varsigma re^{-\ti\varphi}-\zeta^\ast re^{\ti\varphi}+ \ti \ell\varphi}.
\end{equation}
Here, we have already separated the angular and radial integrations, which is possible since we will see that both integrals exist. The integration can be carried out in subsequent steps using the two integrals
\begin{gather}
\label{eq:angular_integral}
\int_0^{2\pi}\td\vartheta\,\te^{a\te^{-\ti\vartheta}}\te^{-b\te^{\ti\vartheta}}\te^{\ti \ell\vartheta}=2\pi \left(\frac{a}{b}\right)^{\frac{\ell}{2}} J_\ell(2\sqrt{ab}),\\
\label{eq:radial_integral}
\int_0^\infty\td x\,x^{\ell+1}L_\jmath^{(\ell)}(b x^2)J_\ell(yx)\te^{-a x^2}=\frac{(a-b)^\jmath y^\ell}{2^{\ell+1}a^{\jmath+\ell+1}}L_\jmath^{(\ell)}\left(\frac{b y^2}{4a[b-a]}\right)\te^{-\frac{y^2}{4a}}.
\end{gather}
The upper one can easily be derived by writing the first two exponentials in their power series and using the series representation of the Bessel functions of the first kind $J_\ell(x)$ (see, for example, 8.440 in Ref.~\cite{Gradshtyn1980}). Once the angular integration is performed, the remaining radial integral can the carried out using the lower integral, which is valid for $\Re\,(a)>0$ and can be found, for example, in 7.421 4. of Ref.~\cite{Gradshtyn1980}. In total, for $I_{\jmath,\ell}$ this yields
\begin{equation}
\label{eq:integral2}
\int\td^2\xi\,\xi^\ell L_\jmath^{(\ell)}\big(q|\xi|^2\big)\te^{-c|\xi|^2}\te^{\varsigma\xi^\ast-\zeta^\ast\xi}
=\frac{\pi(c-q)^\jmath}{c^{\jmath+\ell+1}} \varsigma^\ell L_\jmath^{(\ell)}\left(\frac{q\varsigma\zeta^\ast}{c[q-c]}\right)\te^{-\frac{\varsigma\zeta^\ast}{c}}
\end{equation}
and the case where the monomial in this integral is $\xi^{\ast \ell}$, instead of $\xi^\ell$, is easily derived from this expression by replacing $\varsigma$ with $-\zeta$, and vice versa, and taking the complex conjugate.

For the special case $q=c$ one has to perform the limit
\begin{equation}
\label{eq:laguerre_limit}
\lim_{x\to 0}(-x)^\jmath L_\jmath^{(\ell)}\left(\frac{z}{x}\right)=\frac{z^\jmath}{\jmath!}
\end{equation}
in order to obtain
\begin{equation}
\label{eq:integral3}
\int\td^2\xi\, \xi^\ell L_\jmath^{(\ell)}\big(c|\xi|^2\big)\te^{-c|\xi|^2}\te^{\varsigma\xi^\ast-\zeta^\ast\xi}=\frac{\pi}{\jmath!c^{\jmath+\ell+1}}
\varsigma^\ell (\varsigma\zeta^\ast)^\jmath \te^{-\frac{\varsigma\zeta^\ast}{c}},
\end{equation}
which we used twice in the main text.

\end{document}